\documentstyle[twoside,fleqn,epsfig,espcrc2]{article}
\title{Finite density [{\small might well be easier}] at finite temperature}
\author{Maria-Paola Lombardo \address{Istituto Nazionale di Fisica
Nucleare \\ Laboratori Nazionali del Gran Sasso}}
\begin{document}
\begin{abstract}
{Experiments with  imaginary chemical 
potential and Glasgow method carried out in two interrelated models -- 
four dimensional QCD in the infinite coupling limit,
and one dimensional QCD --  support the point of view expressed by 
the title.}
\end{abstract}
\maketitle

The profound differences of  dense, cold matter, and hot, 
moderately dense, hadron gas  might  be reflected by  different 
performances of  numerical approaches to finite density QCD.
In particular, the enhancement of the fluctuations near
$T_c$ should give better
chances either to the Glasgow method \cite{IMB}, by ameliorating the
overlap problem,  and to the imaginary chemical potential\cite{AKW}, 
which does not
systematically bias the ensemble and thus needs large response to
infinitesimal $\mu$ changes near $\mu=0$. 

The fluctuations of baryons are measured by 
the baryon number susceptibility
\begin{equation}
\chi(T,\mu) = \partial \rho (\mu, T) / \partial \mu 
= \partial^2 log Z (\mu, T) / \partial \mu^2
\end{equation}
Lattice results  \cite {GLRST}  indicate the range where 
$\chi(T,\mu=0)$ is significantly different from zero : this is 
the candidate region for performing finite density calculations  below
$T_c$ -- a narrow, but not minuscule interval. 

As $Z(\mu)$ (we shall explicit only the  $\mu$ dependence)  
is an even function of $\mu$, we have: 
\begin{eqnarray}
log Z (\mu)& = & K + a \mu^2 + b (\mu^4) + O(\mu^6) \\
log Z (\nu)& = & K  -a \nu^2 + b (\nu^4) + O(\nu^6)
\end{eqnarray} 
where $\nu = i \mu$. Note that $a =\chi(T,0) $. Consider
\begin{eqnarray}  
a_{eff} (\mu) & = & (log Z ( \mu) - log Z (\nu))/2\mu^2 \\
b_{eff} (\mu) & = & (log Z ( \mu) + log Z (\nu) - 2K)/2\mu^4
\end{eqnarray}
A  fourth order polynomial describes the data where 
$a_{eff}$ and $b_{eff}$ do not depend on $\mu$.

The strong coupling limit of QCD displays confinement and
chiral symmetry breaking, and, besides having produced interesting,
semiqualitative results on the phase diagram, offers a nice test bed for 
numerical methods.

Consider the effective potential $V_{eff}$ 
(related to $Z$ as $Z = (\int V_{eff} d \lambda)^{V_s}$ ) 
in the $g=\infty$ limit of 
QCD \cite{strong}:
\begin{eqnarray}
V_{eff}(\lambda, \mu) &=&  2cosh(rN_tN_c\mu) + \\ 
&&   sinh[(N_t+1)N_c \lambda] /sinh(N_t \lambda) \nonumber
\end{eqnarray}
where the variational parameter $\lambda$  
(essentially, $<\bar \psi \psi>$)  is to be determined
by a minimum condition, 
$N_t$ is the number of points in time direction,
$N_c$ the number of colors (which we fix: $N_c=3$), 
r the asymmetry factor $a_\tau/a_\beta$,
and $rN_t$ is the inverse temperature. Note
that formally $V_{eff}(\lambda, \mu)$ is
an one dimensional partition function which ``remembers'' its four 
dimensional origin via $\lambda$.

\begin{figure}[t]
\epsfxsize=19pc %
\epsfbox{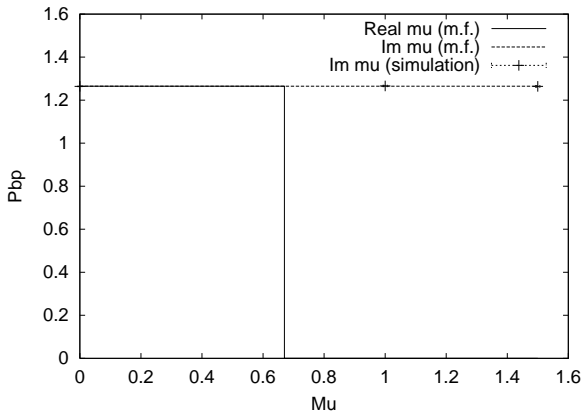} 
\caption{$<\bar \psi \psi >$ as a function of real
 and imaginary chemical potential (a constant) in 
the zero temperature limit.  The only relevant quantity
-- $\mu_c$--  is not amenable to an analytic continuation from
imaginary $\mu$.}
\end{figure}
\begin{figure}[t]
\epsfxsize=19pc %
\epsfbox{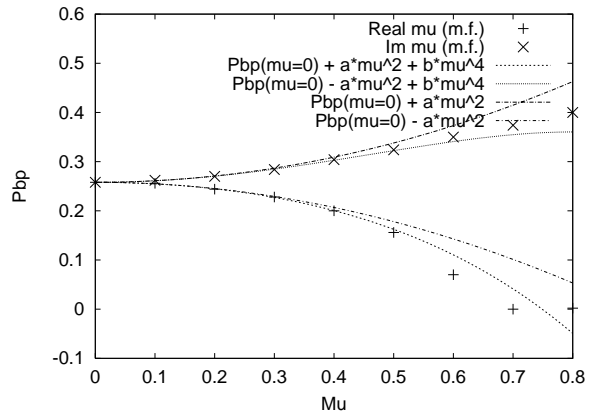} 
\caption{$<\bar \psi \psi >$ as a function of real
(crosses) and imaginary chemical potential (plus'es)  
for $T$ slightly below $T_c$. 
A  polynomial representation (eqs. (2) and (3)) describes well either
data set. $a$ and $b$ are computed as in eqs. (4) and (5).}
\end{figure}

For a purely imaginary chemical potential, 
$cosh(r N_t N_c \mu) \rightarrow cos(r N_t N_c\mu)$. 
We see the expected periodicity $ 2 \pi /(r N_t N_c)$,
and we note that the chemical potential term can be ignored
for large $N_t$ (zero temperature).
 Indeed, we have verified that in
this limit any dependence on the chemical potential is lost  
\cite{fl}, Fig.1 : at $T=0$ 
$<\bar \psi \psi>$ as a function of (complex) $\mu$ is a constant in the
half plane $\Re(\mu)  < \mu_c$, $\mu_c$ being the (real) critical
chemical potential.
By increasing the temperature, the effective potential changes with
imaginary $\mu$, and the chiral condensate {\em increases}  
with imaginary chemical potential, as it should. 
This behaviour is shown in Fig. 2, where we plot the 
chiral condensate as a function of a real, and a purely 
imaginary $\mu$ rather close to $T_c$. 
We note that the fourth order polynomial
behaviour anticipated above holds in a rather large range of chemical
potential, and that a simple second order polinomial works fine as well.

\begin{figure}[t]
\epsfxsize=19pc %
\epsfbox{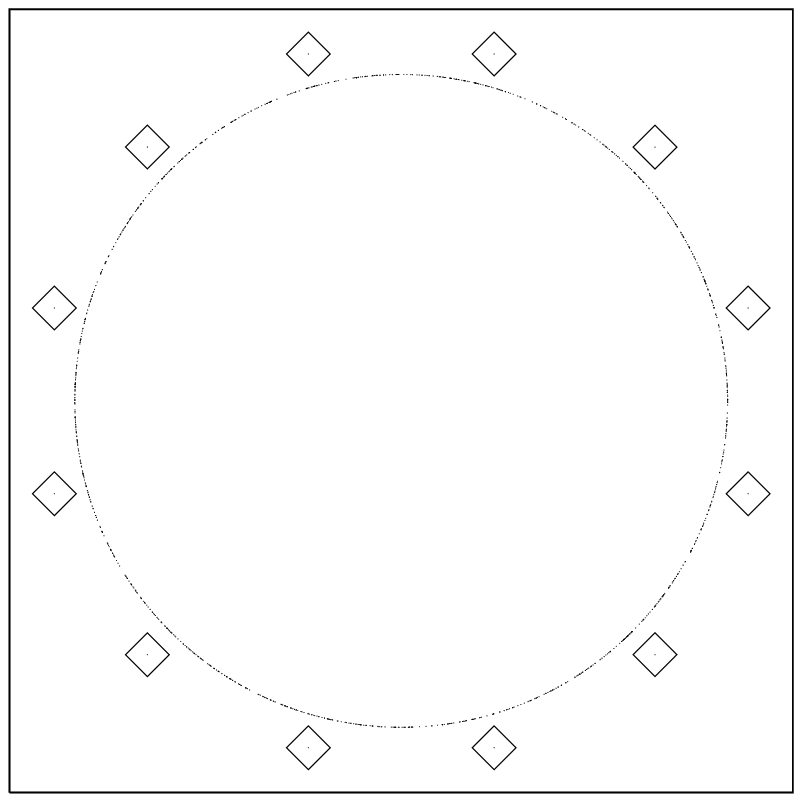}
\caption{Distribution of the zeros of the determinant (100
configurations) and the exact results for the zeros of the partition
function $Z$ with $N_t=4$ (diamonds) for one dimensional QCD. }
\end{figure}
\begin{figure}[t]
\epsfxsize=19pc %
\epsfbox{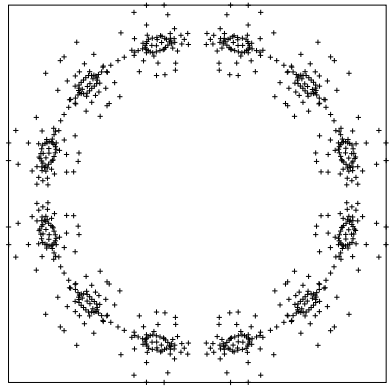} 
\caption{The zeros of $Z = <Det>$ from the Glasgow method approach
the exact pattern Fig. 3. Correspondingly, number density and chiral 
condensate become nearly exact.}
\end{figure}

In the same hot region reweighting too might prove useful. To check that,
we have tested the Glasgow method 
in one dimensional QCD\cite{exact}, a solvable 
model without SSB, but with baryons, whose partition function
is formally identical to the effective potential  above. 
The crucial
difference comes from the meaning of the parameter $\lambda$
which here is simply $sh^{-1} m$. The bare mass $m$ plays the r\^ole
of an explicit breaking term, and $<\bar \psi \psi> \ne 0$
when $m \ne 0$. Fixing $N_c=3$, the relevant parameter
is $N_t \lambda$, which is the analogous of an inverse temperature.

At large ``temperature'' (small $N_t \lambda$) we have verified
that the Glasgow method reproduces the exact results \cite{touka},
thus supporting the idea that reweighting
methods can be successfully used in that regime. We would like
to stress that this is a non trivial results since at large temperature
quenched and exact results are distinct
\cite{exact}. This contrasts
with the low temperature regime of the same model, where the quenched
approximation is apparently exact, and the success of the Glasgow
method is then guaranteed a priori (as the Glasgow method reproduces
the quenched results for small statistics).

Consider the zeros of the determinant, 
which control the singularities of the quenched model,
and those of the partition function, the Lee Yang zeros controlling
the singularities of the full model, in the complex  plane 
$ e^{\mu}$. These $N_t \times N_c$ zeros
can be easily calculated using the results
of \cite{exact} and are shown in Figs. 3. The Glasgow method (Fig. 4) 
reproduces them  correctly, as expected given that $Z$ itself approaches 
the correct value as mentioned above.  The smoothening of the transition in the
full model noted in \cite{exact} can be clearly read off Figs. 3, 4
: the zeros of $Z$ move away from the real axis. It should 
be noticed, however, that  in this case there is no pathological  onset 
(no Goldstone particle here!): the Glasgow method only needs rearranging 
zeros  on circular patterns. It is  reasonable to hope 
that the same happens  in high temperature,  four dimensional QCD.  From
Fig. 3 we also understand why the quenched approximation  is apparently 
working on long lattices \cite{exact}: the zeros will become  dense on the 
same circle where the zeros of the determinant are randomly distributed. An
important difference between the full and quenched 
transition however remains: only the zeros of the
partition functions are fixed by symmetries. 
Hence, only the  nearest zero of Z
(and not that of the determinant) 
has a calculable distance from the real axes $\simeq \mu_c/(3N_t)$,
fulfilling a Lee-Yang--like scaling towards the $N_t = \infty$ 
singularity.

We have given examples of clear  signals obtained by
use of an imaginary chemical potential, and successful calculations
with the Glasgow method in toy models related with QCD. We 
have suggested a possible simple line of analysis
which combines and cross checks results from both approaches.

\section*{Acknowledgements}

I would like to thank John Kogut and Don Sinclair for
discussions  on finite density at high temperature,  
Philippe de Forcrand for initial collaboration 
on imaginary chemical potential, Ian Barbour for kindly
making his codes available, and for discussions. 
This work was partly supported by the NATO Collaborative Research 
Grant  {\em Lattice QCD at Non--zero Temperature and Chemical Potential},
no. 950896,  and by the  TMR network {\em Finite Temperature Phase Transitions 
in Particle Physics}, EU contract no. ERBFMRXCT97-0122.

\end{document}